\documentclass[12pt]{article}
\usepackage{amssymb,amscd}
\usepackage{graphicx}
\usepackage{multirow}
\usepackage{float}
\usepackage{amsmath}
\usepackage{authblk}

\begin{document}

\title{Edge plasma relaxations due to diamagnetic stabilization}
% Force line breaks with \\
\author{F. Cianfrani\footnote{f.cianfrani@fz-juelich.de}}
\author{G. Fuhr} 
\author{P. Beyer}
\affil{%
Aix-Marseille Universit\'e, CNRS, PIIM, UMR 7345 Marseille, France%\\This line break forced% with \\
}%

\date{\today}% It is always \today, today,

\maketitle

\begin{abstract}
A new mechanism for pressure profile relaxations in an edge tokamak plasma is derived from simulations within the two-fluid three-dimensional turbulence code EMEDGE3D. The relaxation is due to diamagnetic coupling in the resistive ballooning/drift wave dynamics: unstable modes experience explosive growth at high pressure gradients after a phase in which they are stabilized by the diamagnetic coupling leading to the onset of a transport barrier. The sheared $E\times B$ flow does not play any significant role. After relaxation the transport barrier forms again and it sets the conditions for a novel relaxation, resulting in an oscillatory behavior. We find that energy flux into the scrape of layer decreases with increasing oscillation frequency and that the oscillations are tamed by increasing plasma temperature. This behavior is reminiscent of so-called type III Edge Localized Modes. A one-dimensional model reproducing the relaxations is also derived.     
\end{abstract}

\section{Introduction}

The control of heat deposition to the wall of forthcoming tokamaks is one of the major issues in fusion research. The H-mode regime \cite{Wagner:1982} realizes the best conditions for reactor operation, with stronger edge pressure gradients and reduced heat deposition to the plasma facing components with respect to the conventional L-mode. The establishment of a steep pressure profile can be understood in terms of the formation of a transport barrier, at which the transport of particles and energy gets reduced. However, the edge of a magnetically confined plasma is not generically stationary, rather it is subject to relaxation oscillations. One important class are the edge localized modes (ELMs) where the edge transport barrier relaxes quasi-periodically, each relaxation leading to the release of energy and particles \cite{Connor:1998,Connor:2000}. The theoretical comprehension and the actual control of ELMs is crucial to avoid wall erosion and loss of confinement, while they could also be beneficial for impurities and ash removal.     

The ELMs have been classified into three main classes \cite{Doyle:1991,Zohm:1996,Suttrop:2000}: type I, II and III. Type-III ELMs occur at lower injected power, edge density and edge temperature with respect to type-I ELMs, but the main distinction among then is with respect to the behavior of the energy flux $Q_{SOL}$ into the scrape of layer. In type-III ELMs the oscillation frequency decreases with increasing $Q_{SOL}$, while in type-I it increases. 

The plasma confinement and dynamics close to the LH transition is known to be dominated by resistive micro-instabilities \cite{Suttrop:2000}. In between relaxations a transport barrier develops, as the dominant instability is either quenched or close to marginal stability, growing up on a slower time scale with respect to the particle and energy diffusion time. Several mechanisms for relaxations have been discussed, involving in toroidal devices the poloidal $E\times B$ shear flow of the plasma (or, equivalently, the radial electric field) that also plays an essential role in the generation of the transport barrier. The shear flow either actively takes part in the dynamics, by quenching turbulence, or induces stabilization of the underlying instability with a time delay. Both mechanisms can lead to relaxation oscillations, see e.g. \cite{Bian:2003} and \cite{Beyer:2005}.

In this work, we show the existence of a different mechanism for relaxation oscillations in which the ExB shear flow is not relevant. This mechanism is based on diamagnetic effects, in particular the electron pressure gradient in Ohm’s law. Diamagnetic effects are known to have several consequences in magnetically confined plasmas: they are at the origin of electrostatic drift waves \cite{Hasegawa:1983}, they have a stabilizing effect on ideal \cite{Rogers:1999,Huysmans:2001} and resistive \cite{Kim:1991} ballooning modes, and they have a stabilizing influence on current driven instabilities e.g. tearing modes \cite{Meshcheriakov:2012}. Also, simulations have shown that diamagnetic effects are sufficient to drive relaxations in the plasma core (so-called sawtooth cycling) \cite{Halpern:2011}. 

In particular, here we demonstrate in 3D electromagnetic turbulence simulations of the plasma edge that diamagnetic effects can drive relaxations of the pressure profile. The simulations are performed through the nonlinear two-fluid code EMEDGE3D. EMEDGE3D has been successfully used to predict the onset of a transport barrier for flux-driven simulations in the presence of a neo-classical friction term both in the electro-static limit \cite{Chone:2014,Chone:2015} and in the full electro-magnetic case \cite{DeDominici:2019}, reproducing qualitatively also the isotope effect on the power threshold for the pedestal formation \cite{DeDominici:2019b}. 

In this work, we focus on the impact of the diamagnetic coupling and we outline its crucial role in predicting a particular class of oscillations, that do not arise when the diamagnetic effects are artificially turned off. The beginning of the relaxations are characterized by some bumps in the main flux contributions, namely the convective flux and the flux due to the non-linear magnetic field perturbations, that are negligible otherwise. The inspection of the fields profiles suggests that the following mechanism is at work during the observed oscillations: because of diamagnetic stabilization the fluctuations are close to marginal stability and they start to be destabilized just when pressure gradients are large, after which their amplitudes increase up to provide a crash of the equilibrium pressure profile. During the relaxation, a net energy flux into the scrape of layer is observed and the average released energy decreases with increasing source flux. Since, the oscillation frequency is roughly proportional to the source flux, it turns out that the energy flux into the scrape of layer decreases with increasing oscillation frequency. This feature and the fact that the oscillations are tamed by increasing plasma temperature provide a similarity with type-III ELMs. %allows us to conclude that the observed oscillations reproduce type-III ELMs. 
Finally, a 1D model is defined that is able to generate the same kind of oscillations observed in 3D simulations.

\section{3D simulations}
EMEDGE3D is based on a fluid description for the outer part of a tokamak plasma, {\it i.e.} for minor radius $r$ larger than a reference value $r_0$. Constant density $n_0$, equal electron and ion temperatures $T_e=T_i=T$ and a constant toroidal external magnetic field $B_0$ (slab approximation) are assumed. The variables are the electronic pressure $p$, the electro-static potential $\phi$ and the parallel (to the external B-field) component of the vector potential $\psi$. Time is normalized through the drift time $\tau=L_p/c_s$, $L_p$ being a conventional fixed length scale and $c_s=\sqrt{T/m_e}$ with the electron mass $m_e$, while the radial coordinate is $x=(r-r_0)/\rho_i$, $\rho_i$ being the ion Larmor radius. The angular coordinates $y$ and $z$ are obtained through a rescaling of the poloidal and toroidal angles, $\theta$ and $\varphi$. By a proper normalization of $p$, $\phi$ and $\psi$ the nonlinear equations for charge neutrality, energy balance and the Ohm's law read as follows \cite{Beyer:1998,Fuhr:2008}  
\begin{eqnarray}
&\partial_t \nabla_\bot^2 \phi + \{\phi,\nabla_\bot^2 \phi\} = - \,\nabla_\parallel\,\nabla_\bot^2\psi - \omega_D \,{\bf G}p + \mu_\bot \nabla_\bot^4 \phi \label{charge}
\\
&\partial_t p + \{\phi,p\} = - \Gamma\,\nabla_\parallel\,\nabla_\bot^2\psi + \delta_c \,{\bf G}(\phi-C_{dia}\,p) +\chi_\bot \nabla_\bot^2 p + S  \label{energy}
\\
&\beta_e\partial_t \psi = -\epsilon_p^2\nabla_\parallel(\phi-C_{dia}\,p) + \eta\,\beta_e\, \nabla_\bot^2 \psi\,,
\label{Ohm}
\end{eqnarray}
where $\nabla_\bot^2=\partial_x^2+\partial_y^2$ and $\nabla_\parallel ..= \partial_\parallel ..- (\beta_e/\epsilon_p^2)\,\{\psi,..\}$, with $\partial_\parallel$ the derivative along the B-field
\begin{equation} 
\partial_\parallel=\left(\partial_z+\frac{a}{q\rho_i}\,\partial_y\right)\,,
\end{equation}
$a$ being the minor tokamak radius and $q$ is the safety factor. The Poisson brackets $\{..\}$ are defined as $\{f,g\}= \partial_xf\,\partial_y g -\partial_y f\, \partial_x g$ and $G$ denotes the toroidal curvature operator 
\begin{equation}
{\bf G}= \cos\theta\,\partial_y + \sin\theta\,\partial_x\,, 
\end{equation}
whose corresponding parameters read $\omega_D=2\,L_p/R_0 $ and $\delta_c=\Gamma\, \omega_D$, with $\Gamma=5/3$ the adiabatic index and $R_0$ the major tokamak radius. The parameter $\epsilon_p$ equals $L_p/R_0$, while other parameters are the dissipative ones, namely viscosity $\nu$, resistivity $\eta$ and perpendicular diffusivity $\chi_\bot$. The plasma beta parameter $\beta_e=p_0/(4\pi B_0^2)$ is computed from a nominal pressure value $p_0=n_0 T_0$ which also enters $p$ normalization. The last factor $C_{dia}$ is considered in order to turn on ($C_{dia}=1$) or to turn off ($C_{dia}=0$) the diamagnetic contribution.
 
The proper simulation region is a toroidal shell and is bounded both internally and externally by a buffer region in which diffusivity is artificially increased to prevent turbulence development there. In the internal buffer region it is placed the power source term $S$ that is constant in the toroidal and polodial directions and has a Gaussian shape in the radial direction $x$. It simulates the heat deposition from the core plasma region and it is characterized by the corresponding total energy flux $Q$
\begin{equation}
Q=\int dx\, S\,.
\end{equation} 

\begin{figure}[H]
\centering
  \begin{tabular}{cc}
    \includegraphics[width=.4\textwidth]{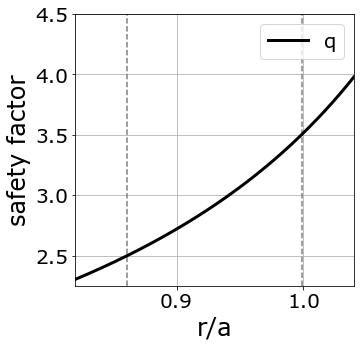} &
		\includegraphics[width=.4\textwidth]{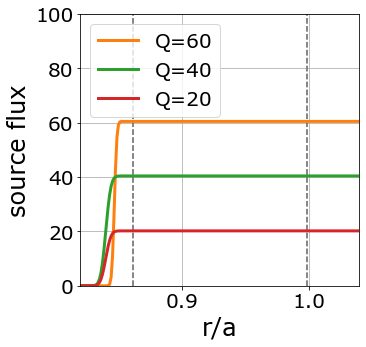}
	\end{tabular}
  \caption{The $q$-profile (left) and the source fluxes (right) for $Q=60$ (orange), $Q=40$ (green) and $Q=20$ (red). The source is placed in the inner buffer region, {\it i.e.} to the left of the first dashed line.}
\label{fig_q_flux}
\end{figure}

In the simulations we consider a medium-sized plasma with major radius $R_0=1.75\,m$, minor radius $a=0.45\,m$ and magnetic field $B_0= 1\,T$. Furthermore, the nominal density and temperature are fixed as $n_0=2.5\times 10^{19}\, m^{-3} $ and $T_0= 50\,keV$, resulting in $\beta_e=2.5\times 10^{-4}$. The safety factor $q$ has a hyperbolic profile ranging from $2.5$ to $3.5$ within the simulation region and it is shown in figure \ref{fig_q_flux} together with the considered source fluxes ($Q=60,40,20$). The radial scale is discretized over 192 points, covering the simulation region ($0.86<r/a<1$), which is bounded by the two dashed lines in figure \ref{fig_q_flux}, and both the buffer regions ($r_{min}/a=0.82<r/a<0.86$ and $1<r/a<r_{max}/a=1.04$). The spectral domain in poloidal and toroidal wave numbers extends from $(m,n)=(0,0)$ up to $(m,n)=(129,32)$ with $\Delta n= 4$ and $\Delta m=1$. The dissipative parameters in the simulation region take the following values in physical units.%\footnote{Indeed, $\chi_\bot$ used in the simulations is taken smaller that the actual value measured in experiments $\sim 1 m^2/s$ \cite{Gruber:1999} in order to prevent the development of instabilities.}
\begin{equation}
\nu=2.4\times 10^{-8}\, kg/ms\qquad \chi_\bot=0.28\, m^2/s \qquad \eta= 2.1\times 10^{-6} \,\Omega\, m\,.
\end{equation}
The simulations are performed from an initial state in which only random-phase fluctuations are present. 

\subsection{Average pressure}
Starting with zero pressure, the increase and the subsequent saturation of the pressure profile can be observed through the value of spatially average pressure 
\begin{equation}
\langle p \rangle =\frac{1}{(2\pi)^2(r_{max}-r_{min})} \int_{r_{min}}^{r_{max}} dr \int d\theta d\varphi \,p\,, 
\end{equation}
and scanned with respect to the total source energy flux $Q$, the $\beta_e$ factor and the parameter $C_{dia}=0,1$ turning off and on the diamagnetic terms. An increase of $\langle p \rangle$ generically implies steepening of pressure gradients. 

\begin{figure}%[H]
\centering
    \includegraphics[width=.75\textwidth]{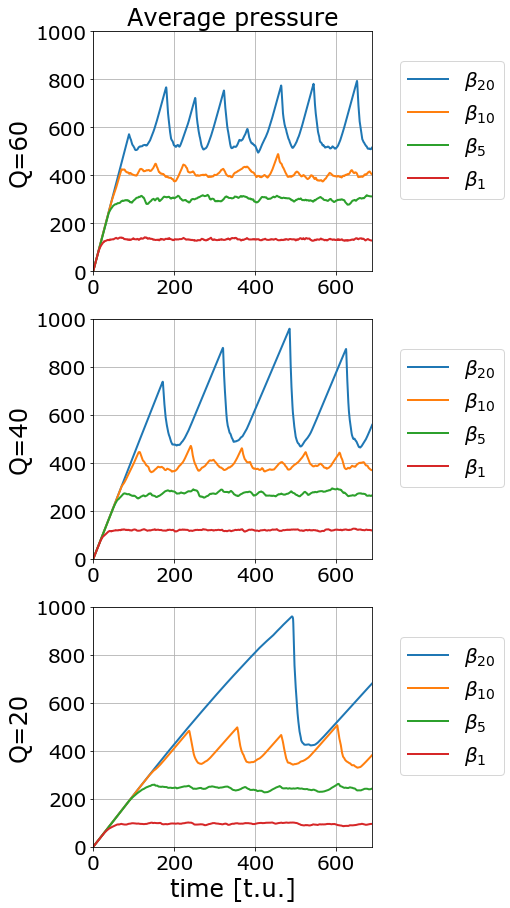} 
  \caption{Average pressure vs time for $\beta_d=\beta_e/d$ with $\beta_e=2.5\cdot 10^{-4}$ and $C_{dia}=1$. The plots are for $Q=60$ (top), $Q=40$ (center) and $Q=20$ (bottom).}
	\label{fig1}
\end{figure}

The results are presented in figure \ref{fig1} for cases including the diamagnetic term. Time is expressed in normalized time units (t.u.) with $100\,t.u.= 0.61\,ms$. 
  
It is worth noting how the average pressure is significantly influenced by the value of the factor $\beta_d$. For $\beta_d= \beta_e$ or $\beta_d=\beta_e/5$, it firstly grows up almost linearly and then reaches a saturated level. This is the behavior of a quiescent plasma with an almost stationary heat deposition to the wall. In fact, the pressure profile and gradients initially grow up due to linear diffusion of the injected power and the fluctuations are destabilized by the onset of linearly unstable modes. The pressure gradient leads to two kinds of unstable modes here: drift waves, driven by the diamagnetic coupling, and resistive ballooning modes, which are destabilized by the curvature. As the fluctuations amplitude increases, non-linear interactions start to dissipate the injected power, up to the saturated point at which the power influx is balanced by nonlinear dissipation and the pressure profile stabilizes. 

The bursty behavior is that observed for $\beta_d=\beta_e/20$ and partially for $\beta_d= \beta_e/10$. In fact, the initial phase during which $\langle p \rangle$ grows linearly is now much longer and it suddenly ends through a rapid relaxation. Then, a new linear phase starts ended again through a rapid relaxation and so on, resembling sawthoot oscillations. 

\begin{figure}%[H]
\centering
  \begin{tabular}{cc}
%	{{\large Source/$\beta$}}	
  \includegraphics[width=.45\textwidth]{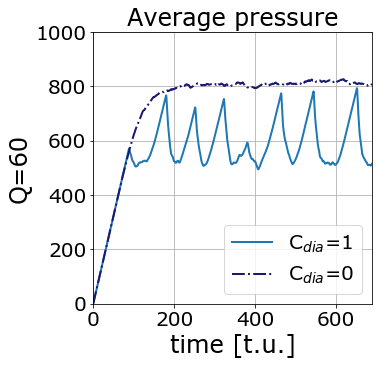} & \includegraphics[width=.45\textwidth]{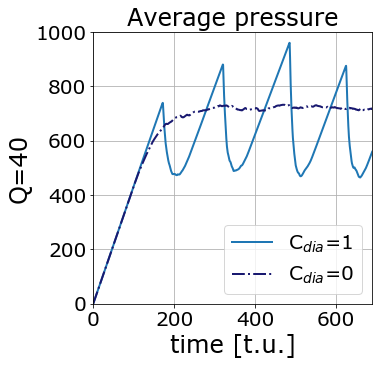} 
		\end{tabular}
  \caption{Average pressure vs time with (solid lines) and without (dash-dotted lines) the diamagnetic term for $\beta_d=\beta_e/20$, $Q=60$ (left) and $Q=40$ (right).}
\label{fig2}
\end{figure}

It is worth noting the comparison with the case in which the diamagnetic contribution is turned off ($C_{dia}=0$), shown in figure \ref{fig2}, which instead exhibits the quiescent behavior observed for larger $\beta_e$'s. Therefore, the bursty behavior, {\it i.e.} the alternation of linear growth of $\langle p\rangle$ and rapid relaxation, is due to the presence of the diamagnetic coupling in Eqs. (\ref{energy}) and (\ref{Ohm}).

\subsection{Equilibrium and fluctuation profiles}
\begin{figure}%[H]
\centering
    \includegraphics[width=.8\textwidth]{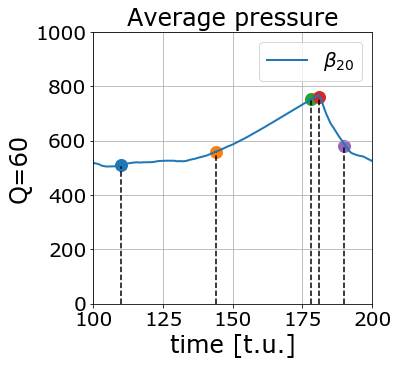} 
  \caption{Zoom of the plot in figure \ref{fig1} for $\beta_d=\beta_e/20$ and $Q=60$ around the peak at $180\,t.u.$.}
\label{fig_1osc}
\end{figure}

\begin{figure}%[H]
\centering
    \includegraphics[width=.6\textwidth]{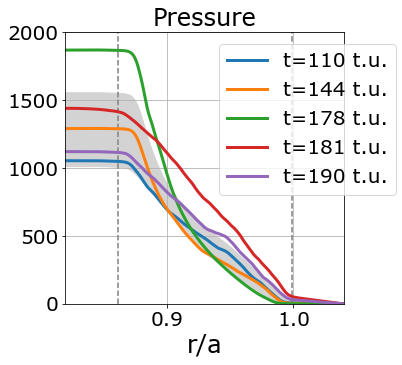} 
  \caption{Pressure profiles before and during the pressure growing phase (blue and orange), just before the top (green) and during the relaxation (red and purple). The curves correspond to the blue, orange, green, red and purple points in the average pressure plot for $Q=60$ and $\beta_d=\beta_e/20$ in figure \ref{fig_1osc}. The grey region is one standard deviation away from the time average pressure.}
\label{fig_p}
\end{figure}

\begin{figure}%[H]
\centering
    \includegraphics[width=.6\textwidth]{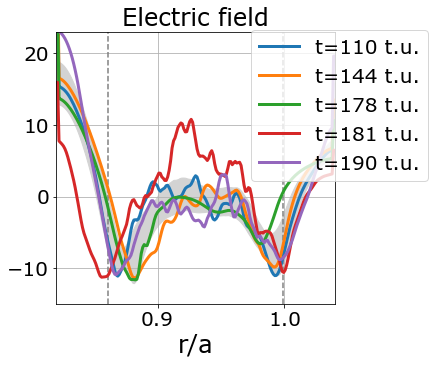} 
  \caption{Electric field profiles before and during the  pressure growing phase (blue and orange), just before the top (green) and during the relaxation (red and purple). The curves correspond to the blue, orange, green, red and purple points in the $\langle p \rangle$ plot for $Q=60$ and $\beta_d=\beta_e/20$ in figure \ref{fig_1osc}. The grey region is one standard deviation away from the time average electric field.}
\label{fig_e}
\end{figure}

In figure \ref{fig_1osc} it is shown one single oscillation with five points outlined for the analysis of the pressure $p_{00}$ and radial electric field $E_r=-\partial_r \phi_{00}$ profiles in figures \ref{fig_p} and \ref{fig_e}, where $p_{00}$ and $\phi_{00}$ are obtained through an average along the angles $\theta$ and $\varphi$, {\it i.e.} $p_{00}=(1/2\pi)^2\int d\theta d\varphi \,p$. 

The blue and orange curves correspond to the phases before and during the onset of the pressure growing phase, the green curves are just before the local maximum of $\langle p \rangle$, while the red and purple ones refer to the beginning and the end of the relaxation. The blue, orange and green curves outline how pressure increases in the inner buffer, where the source is placed and linear diffusion is larger, and the pressure profile steepens signaling the presence of a transport barrier located in the middle of the simulation region. Instead, from the red and purple curves one sees how during relaxation the pressure profile crashes. 

\begin{figure}%[H]
\centering
    \includegraphics[width=.6\textwidth]{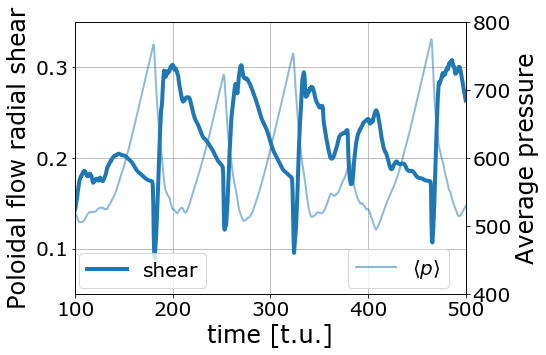} 
  \caption{The shear of the radial electric field (left axis) and the average pressure (right-axis) vs time for $Q=60$. The shear has a sudden decreases just at the beginning of the relaxation, then it has a bump and it decreases again during the relaxation and the pressure growing phase.}
\label{fig_shear}
\end{figure}

Relaxation mechanisms are know \cite{Bian:2003,Beyer:2005} where the $E\times B$ shear plays an important role, either by playing an active part in the (predator-pray like) dynamics or by a constant shear imposing a time delay in fluctuation stabilization. We are not in the second case here, since the shear is not constant, as can be deduced from \ref{fig_e} and even better from figure \ref{fig_shear} in which the average shear is plotted for several oscillations. 

In section \ref{1D} we will see how it is possible to reproduce the relaxations in a 1D model retaining only potential fluctuations and neglecting $\phi_{00}$ and the average $E\times B$ shear as well. This finding rules out also the first case, {\it i.e.} possibility that the $E\times B$ shear plays an active role in the relaxations. Hence, the sheared $E\times B$ poloidal flow is not the physical mechanism behind the relaxations we have found here.

\begin{figure}%[H]
\centering
\begin{tabular}{ccc}
\includegraphics[width=.4\textwidth]{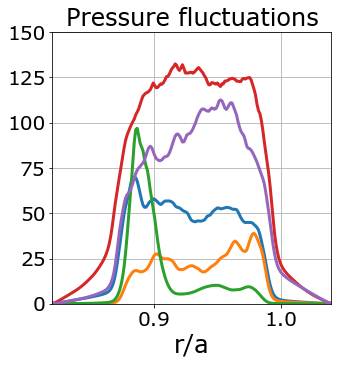} &
\includegraphics[width=.4\textwidth]{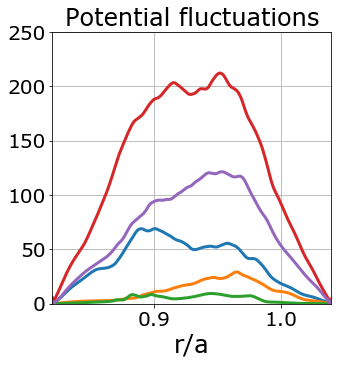} &
\includegraphics[width=.4\textwidth]{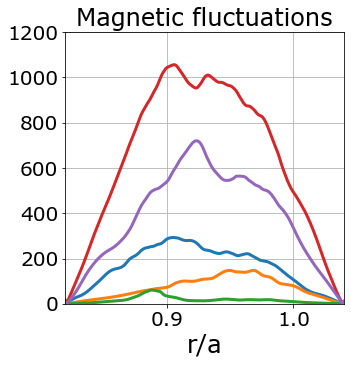} 
\end{tabular}
  \caption{Amplitude of the pressure (left), electro-static potential (center) and $\psi$ (right) fluctuations before and during the pressure growing phase (blue and orange), at just befire the top (green) and during the relaxation (red and purple). The curves correspond to the blue, orange, green, red and purple points in the $\langle p \rangle$ plot for $Q=60$ and $\beta_d=\beta_e/20$ in figure \ref{fig_1osc}.}
\label{fig_flu}
\end{figure}

From figure \ref{fig1} it also turns out that the rapidity of the growth phase is roughly proportional to the source total energy flux $Q$, while the time scale of the relaxation does not significantly depend on $Q$. These facts suggest that during the linear growth the role of fluctuations is negligible and the development of the transport barrier is determined solely by the source energy flux and linear diffusion. 
The plots in figure \ref{fig_flu} confirm this expectation: the amplitude of the fluctuations decreases from the blue to the green curve, they are maximal at the beginning of the relaxation (red curves) and after that they decrease again (purple curves). 

Furthermore, the value of $\beta_d$ determines not only the qualitative behavior, whether it is quiescent or bursty, but also the maximal achievable $\langle p\rangle$. In fact, by increasing $\beta_d$ the average pressure decreases, meaning that pressure gradients are lower and there is more efficient transport to the wall.

However, lowering $\beta_d$ is not enough to have the bursty behavior, since a crucial role is played by the diamagnetic coupling, which quenches the fluctuations, reduce turbulence and set the conditions for the pressure growing phase. The way this mechanism works can be seen in the electro-static approximation: by neglecting the left-hand side of Eq.(\ref{Ohm}), the parallel current $J=\nabla^2_\bot \psi$ can be written in terms of the parallel derivative of $\phi-C_{dia}\,p$ as follows 
\begin{equation}
J=\frac{1}{\eta\beta_e\epsilon_\parallel^2}\nabla_\parallel(\phi-C_{dia}\,p)\,,
\end{equation}
 which can be substituted into Eq.(\ref{energy}) so providing an effective parallel diffusive term $\nabla_\parallel^2 p$ for pressure 
\begin{equation}
\partial_t p + \{\phi,p\} = - \frac{Q}{\eta\beta_e\epsilon_\parallel^2}\nabla^2_\parallel(\phi-C_{dia}\,p) + \delta_c \,{\bf G}(\phi-C_{dia}\,p) +\chi_\bot \nabla_\bot^2 p + S\,.
\end{equation}
%This mechanism, which is present only if the diamagnetic contribution is turned on, acts by suppressing the non-resonating fluctuations ($nq\neq m$) while the resonance condition $nq=m$ holds only in a narrow region because of the magnetic shear also the resonating ones ($nq=m$) so reducing turbulence transport and generating a transport barrier. 
This mechanism is present only if the diamagnetic contribution is turned on and it acts by suppressing the fluctuations having $nq-m\neq 0$. The resonance condition $nq=m$ can hold only in a narrow radial region because of the magnetic shear, so all the fluctuations are generically suppressed resulting in a reduction of the turbulent transport and in the onset of a transport barrier. The performed simulations are electromagnetic, thus we cannot completely rely on the electro-static approximation, but we still obtain a similar stabilization mechanism.
 
\subsection{Fluxes}

Further information is obtained by the analysis of the fluxes governing the dynamics of the pressure profile $p_{00}$, following the equation  
\begin{equation}
\partial_t p_{00} = -\partial_xQ_{tot}+ S\,,\label{fluxes}
\end{equation} 
where the flux $Q_{tot}=Q_{conv}+Q_{\delta B}+Q_{diff}+Q_{curv}$ is the sum of four contributions, the convective, $\delta B$, diffusive and curvature fluxes\footnote{A dissipative contribution due to the curvature term is avoided since it results to be negligible.}, whose explicit expressions read
\begin{align}
&Q_{conv}= \displaystyle\frac{1}{2(2\pi)^2}\int d\theta d\varphi \left(\phi\partial_y p-p\partial_y \phi\right)\\
&Q_{\delta B}=  \displaystyle\frac{Q\beta_e}{2\epsilon_p^2(2\pi)^2}\,\int d\theta d\varphi \left(\nabla_\bot^2\psi\partial_y \psi-\psi\partial_y \nabla_\bot^2\psi\right)\\
&Q_{diff}=-\chi_{\bot}\,\partial_x p_{00}\\
&Q_{curv}=-\displaystyle\frac{\delta_c}{(2\pi)^2}\int d\theta d\varphi \sin\theta \,(\phi-C_{dia}p)\,.
\end{align} 

\begin{figure}%[H]
\centering
\begin{tabular}{c}
\includegraphics[width=.8\textwidth]{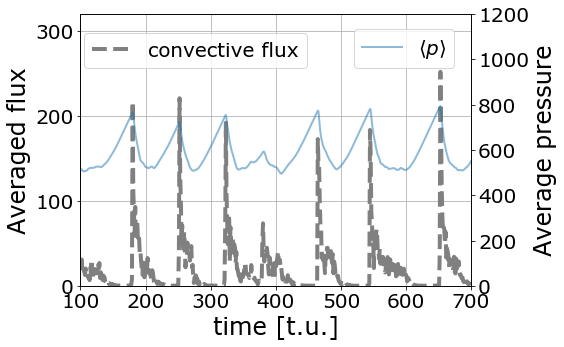} \\
\includegraphics[width=.8\textwidth]{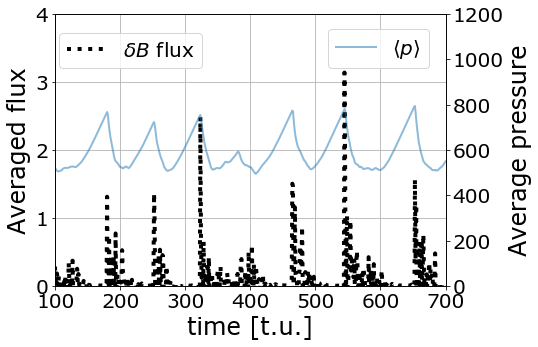} 
\end{tabular}
  \caption{Average convective (top) and $\delta B$ (bottom) fluxes vs time for $Q=60$ and $\beta_d=\beta_e/20$.}
\label{fig3}
\end{figure}

\begin{figure}%[H]
\centering
  \begin{tabular}{c}
%	{{\large Source/$\beta$}}	
  \includegraphics[width=.8\textwidth]{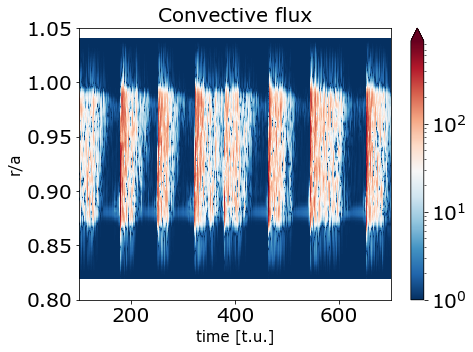} \\ \includegraphics[width=.8\textwidth]{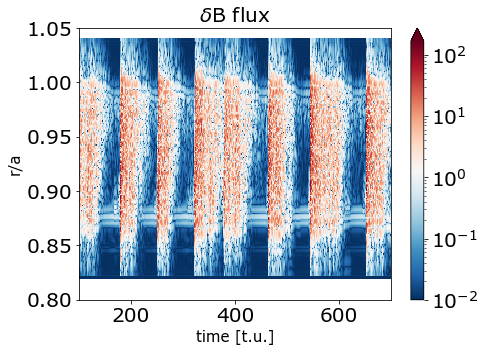} 
		\end{tabular}
  \caption{2D plots of the convective (top) and $\delta B$ (bottom) fluxes with respect to time and minor radius for $Q=60$ and $\beta_d=\beta_e/20$.}
\label{fig4}
\end{figure}

The quench of the fluctuations, thus of the non-linear contributions too, is confirmed by the results shown in figures \ref{fig3} and \ref{fig4} of the convective and $\delta B$ fluxes, which are negligible during the linear growth phase and have a bump only in correspondence to the relaxation phase.    

\begin{figure}%[H]
\centering
\includegraphics[width=.8\textwidth]{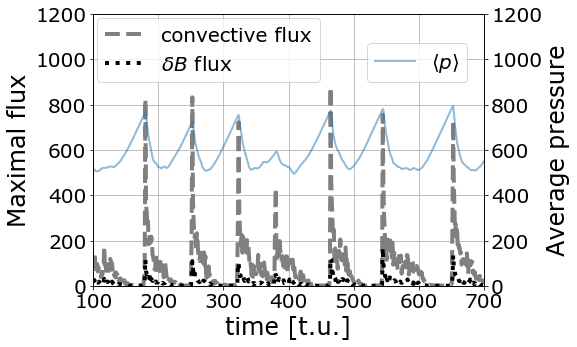} \\
  \caption{Maximal convective (grey) and $\delta B$ (black) fluxes vs time for $Q=60$ and $\beta_e=\beta/20$.}
\label{fig_maxf}
\end{figure}

This bump points out how the fluctuations are finally able to overcome the diamagnetic stabilization mechanism and increase up to back-react on the average pressure profile, producing a sudden decrease of the inner pressure and destroying the transport barrier. Note that the average $\delta B$ flux is a factor 100 lower with respect to the convective flux. However, from 2D plots one sees how the $\delta B$ flux has stronger radial variations and its peaks are at normalized amplitudes of $10^{2}$, thus more or less at the same order of magnitude as those of the convective flux. This is shown in figure \ref{fig_maxf}, in which the maximal fluxes at each time are compared and the $\delta B$ maximal flux is shown to be just a factor 5 smaller with respect to the convective one. 

\begin{figure}%[H]
\centering
\includegraphics[width=.8\textwidth]{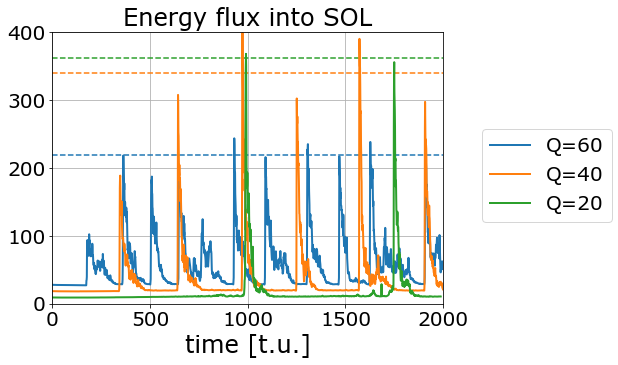} \\
  \caption{The energy flux into the scrape of layer vs time is plotted for $Q=60$ (blue), $Q=40$ (orange) and $Q=20$ (green). The dashed horizontal lines correspond to the average value of $Q_{SOL}$ at the peaks.}
\label{fig_Psol}
\end{figure}

The energy flux into the scrape of layer $Q_{SOL}$ can be defined by integrating Eq.(\ref{fluxes}) along the radial direction $x$. Since the energy flux crossing the inner boundary vanishes identically\footnote{This does not mean that there is no energy flux from the inner plasma region, as it has been modeled through the source flux $Q$ which is placed within the inner buffer region and not on the inner boundary.}, it results 
\begin{equation}
Q_{SOL}= Q -\frac{d\langle p \rangle}{dt}\,,
\end{equation} 
and the corresponding plot is shown in figure \ref{fig_Psol}.  

One observes that the energy flux into the scrape of layer has bumps during relaxations. The dashed lines correspond to the average peaks of $Q_{SOL}$, which decrease with increasing source flux $Q$. Therefore, the peaks of $Q_{SOL}$ on average decrease with increasing frequency, which is the expected behavior for type-III ELMs \cite{Zohm:1996}. Such a correspondence is also supported by the behavior of the average pressure with respect to $\beta_d$ for each $Q$-value in figure \ref{fig1}, which implies that the oscillations are tamed by increasing plasma temperature.

\subsection{Poloidal sections of the fields at the beginning of the relaxation}
\begin{figure}%[H]
\centering
\includegraphics[width=.7\textwidth]{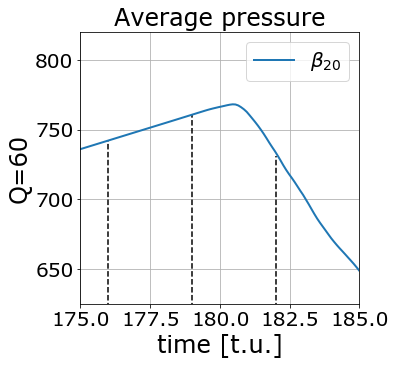} 
  \caption{Zoom of the average pressure vs time plot in figure \ref{fig1} for $Q=60$ and $\beta_d=\beta_e/20$ around the peak at $t=180\,t.u.$. The three dashed vertical lines correspond to the three time values of the 2D plots shown in figures \ref{fig_zoom1}, \ref{fig_zoom2}, \ref{fig_zoom3} and \ref{fig_zoom4}.}
\label{fig_zoom}
\end{figure}

\begin{figure}%[H]
\centering
\includegraphics[width=1.3\textwidth]{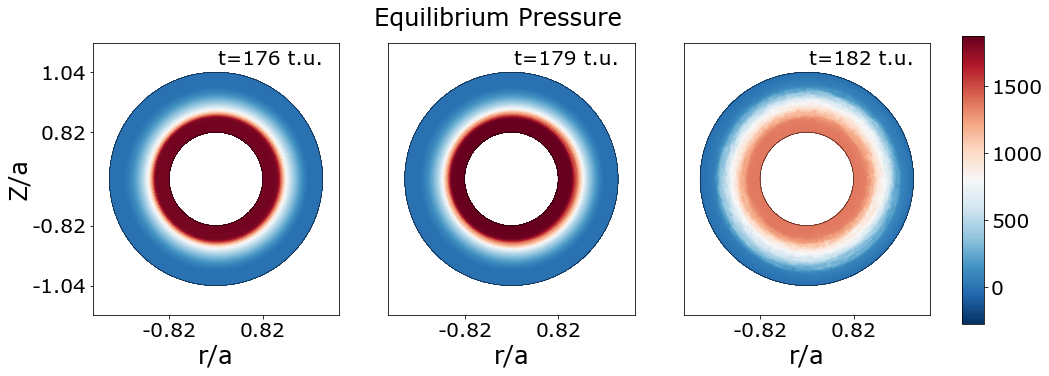} 
  \caption{The poloidal sections of equilibrium pressure (toroidal number $n=0$)  at the three times outlined in figure \ref{fig_zoom}.}
\label{fig_zoom1}
\end{figure}

\begin{figure}%[H]
\centering
\includegraphics[width=1.3\textwidth]{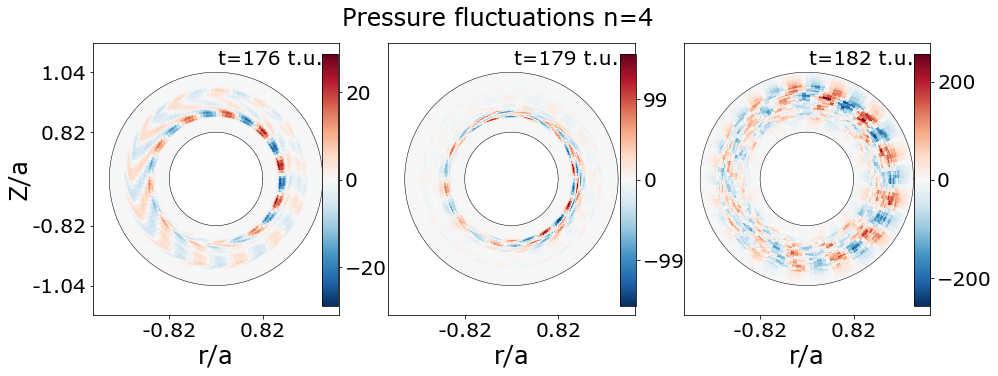} 
  \caption{The poloidal sections of pressure fluctuations (toroidal number $n=4$) at the three times outlined in figure \ref{fig_zoom}.}
\label{fig_zoom2}
\end{figure}

\begin{figure}%[H]
\centering
\includegraphics[width=1.3\textwidth]{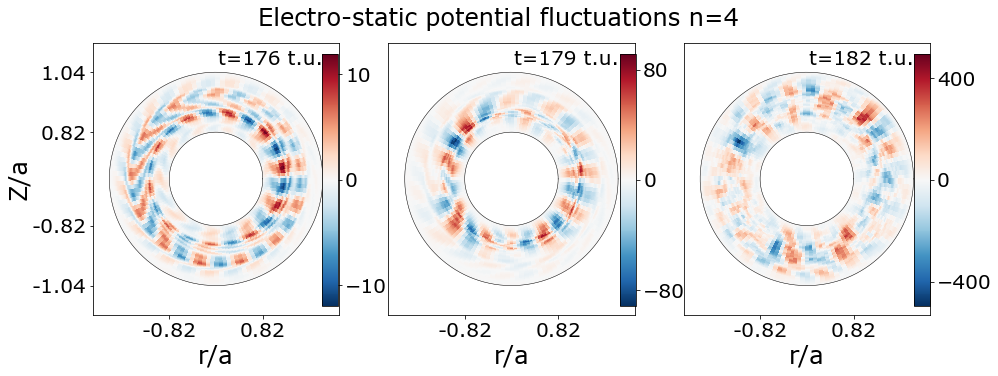} 
  \caption{The poloidal sections of electro-static potential fluctuations (toroidal number $n=4$) at the three times outlined in figure \ref{fig_zoom}.}
\label{fig_zoom3}
\end{figure}

\begin{figure}%[H]
\centering
\includegraphics[width=1.3\textwidth]{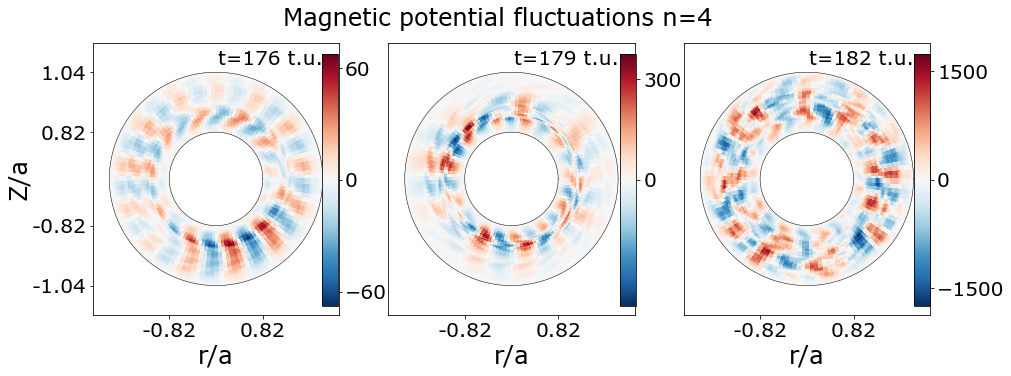} 
  \caption{The poloidal sections of $\psi$ fluctuations (toroidal number $n=4$) at the three times outlined in figure \ref{fig_zoom}.}
\label{fig_zoom4}
\end{figure}

A zoom of the time evolution of the average pressure for $Q=60$ and $\beta_d=\beta_e/20$ is shown in figure \ref{fig_zoom}. The corresponding equilibrium pressure, pressure, electro-static potential and $\psi$ fluctuations are plotted in figures \ref{fig_zoom1}, \ref{fig_zoom2}, \ref{fig_zoom3} and \ref{fig_zoom4}, respectively, for the toroidal wave number $n=4$  (similar plots are obtained for different values of $n$). The buffer and the simulation regions have been amplified for illustrative purposes.

It is worth noting how a single mode dominates the dynamics of pressure fluctuations before the relaxation (left plot in figure \ref{fig_zoom2}), while a turbulent regime is established during relaxation (right plots in \ref{fig_zoom2}, \ref{fig_zoom3} and \ref{fig_zoom4}).% Later, pressure gradients decrease and the system drops below marginal stability so that the fluctuations, and turbulence as well, are suppressed and the pressure profile restarts steepening.

\begin{figure}%[H]
\centering
\includegraphics[width=1.2\textwidth]{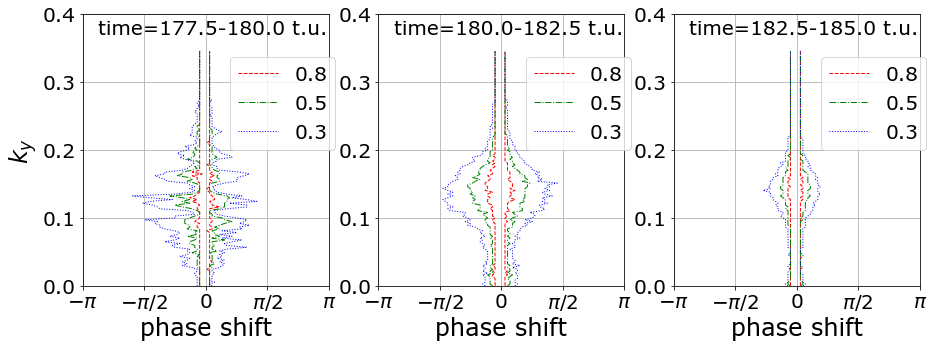} 
  \caption{The wave-number/phase-shift plot for pressure and electro-static potential fluctuations at three sample time intervals before (left) and during (center and right) the relaxation. At fixed $k_y$ the contour plots correspond to the values $0.8$ (red), $0.5$ (green) and $0.3$ (blue) of the phase-shift distribution.}
\label{fig_ky_phase}
\end{figure}

We therefore observe a phase where all fluctuations are suppressed except for one mode. As this phenomenon is not observed in the control case without diamagnetic coupling, we conclude that this coupling is responsible here for the supression of the fluctuations and for the subsequent steepening of the pressure profile in this phase. This phase ends only after the pressure gradients are so large that fluctuations become unstable, reach large amplitudes and back-react on the profile making it drop. The relaxation phase ends when pressure gradients decrease below marginal stability and fluctuations get suppressed again. The fluctuations here are mostly drift waves, as outline by the wave-number/phase-shift plots in figure \ref{fig_ky_phase} showing that the phase-shift for pressure and electro-static potential fluctuations has a narrow distribution close to zero.% \cite{Scott:05}. 

In order to confirm this scenario, a 1D model reproducing the main features of 3D simulations is discussed in the next section. 

\section{1D model}\label{1D}

For constructing the 1D-model, it is sufficient to assume vanishing mean profiles of the electrostatic and magnetic potentials $\phi$ and $\psi$ and to consider a single fluctuating mode $(m,n)$. Hence, the variables are written as follows 
\begin{eqnarray}
&p= \bar{p}(x,t) +\left(e^{i(m\theta-n\varphi)}\tilde{p}(x,t)+\,c.c.\right)\\ 
&\phi= e^{i(m\theta-n\varphi)}\tilde{\phi}(x,t)+\, c.c. \\ 
&\psi= e^{i(m\theta-n\varphi)}\tilde{\psi}(x,t)+\,c.c.\,,\nonumber
\end{eqnarray}
and the system of equations for $\bar{p}$, $\tilde{\phi}$, $\tilde{p}$ and $\tilde{\psi}$ reads 
\begin{equation}
\partial_t \partial_x^2 \tilde\phi -k^2\partial_t \tilde\phi  =  i\,\left(n-\frac{m}{q}\right)\,\nabla_\bot^2\tilde\psi -ik \omega_D\tilde{p} + \mu_\bot (\partial_x^2-k^2)^2 \tilde\phi \label{1Dcharge}
\end{equation}
\begin{equation}
\partial_t \bar{p} = ik\partial_x (\tilde\phi\,\tilde{p}^*-\tilde{p}\,\tilde\phi^*) + ik\,\Gamma\,\beta_e\,\epsilon_\parallel^2\,\partial_x(\tilde\psi^*\,\nabla_\bot^2\tilde\psi-\tilde\psi\,\nabla_\bot^2\tilde\psi^*) +\chi_\bot \partial_x^2 \bar{p} + S\label{1Denergy}  
\end{equation}
\begin{equation}
\partial_t \tilde{p} = ik\tilde\phi\,\partial_x\bar{p} +i \Gamma\,\left(n-\frac{m}{q}\right)\,\nabla_\bot^2\tilde\psi-ik \delta_c \,\left(C_{dia}\tilde{p} - \tilde{\phi}\right) +\chi_\bot (\partial_x^2-k^2)\tilde{p}\label{1Denergytilde}
\end{equation}
\begin{equation}
\beta_e\partial_t \tilde\psi = -\displaystyle\frac{i\alpha}{\epsilon_\parallel^2}\,\left(n-\frac{m}{q}\right)\,\left(C_{dia}\tilde{p}-\tilde{\phi}\right)+ik\,C_{dia}\,\beta_e\,\tilde\psi\,\partial_x\bar{p}+ \eta\,\beta_e\, (\partial_x^2-k^2) \tilde\psi 
\label{1DOhm}\,,
\end{equation}
$k$ being the wave number in $y$ and the curvature operator is the slab one ${\bf G} \sim \partial_y$. 

The 1D model above is solved for $\beta_d=\beta_e/20$  by increasing dissipation (diffusivity, viscosity and resistivity) for the fluctuating components by a factor $50$. Moreover, we artificially amplify the stabilization effect of the diamagnetic term by increasing the parameter $\alpha$ in Eq. (\ref{1DOhm}) from its natural value $\alpha=1$ to $\alpha=2$. These amplifications are necessary because the stabilizing effect due to the non-linear interaction with the other modes (with higher $m$, $n$) is missing and, as we will see, the 1D model results to be less stable compared to 3D simulations. 

\begin{figure}%[H]
\centering
\begin{tabular}{c}
\includegraphics[width=.8\textwidth]{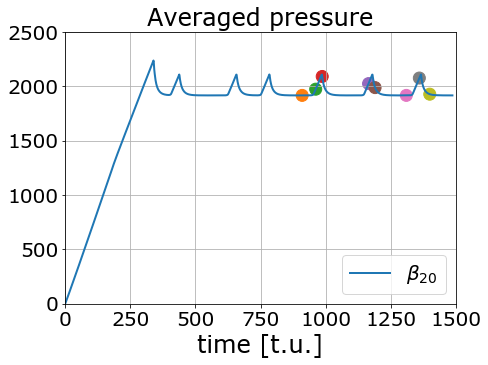} \\
\includegraphics[width=.8\textwidth]{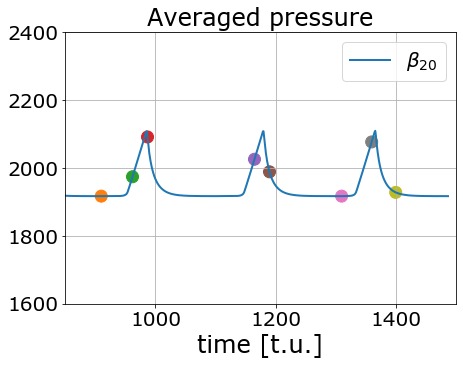} 
\end{tabular}
  \caption{Averaged pressure vs time for the 1D model, $Q=60$ and $C_{dia}=1$. The bottom figure is a zoom of the top one over the last three oscillations.}
%  \caption{A resume of the 1D model results: the $p_{00}$ energy (top left), the pressure profile (top center), the $q$ profile and the position of the resonant surface (top right), the profiles of pressure (bottom left), electro-static potential (bottom center) and magnetic potential (bottom right) fluctuations. The profiles are taken at some selected times, corresponding to the points outlined in the $p_{00}$ energy plot (top left).}
\label{fig_1D2}
\end{figure}

\begin{figure}%[H]
\centering
\includegraphics[width=.8\textwidth]{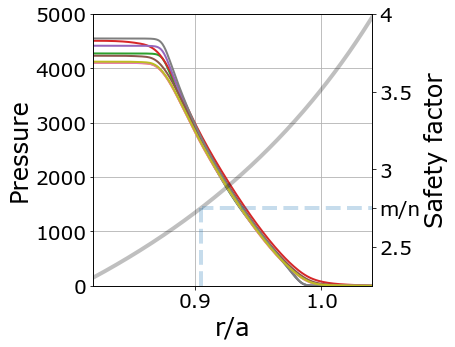} 
  \caption{Pressure profiles (left vertical axis) before and during the pressure growing phase (orange, pink, green and purple), at the top (grey) and during the relaxation (red, brown and yellow). The curves correspond to the orange, green, red, purple, brown, pink, grey and yellow points in figure \ref{fig_1D2}. The safety factor profile (right vertical axis) is plotted in solid grey line and the position of the resonating surface for the chosen mode is outlined through the dashed lines.}
\label{fig_1D_prq}
\end{figure}

\begin{figure}%[H]
\centering
\begin{tabular}{ccc}
\includegraphics[width=.4\textwidth]{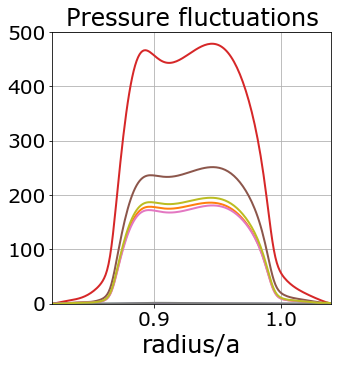} &
\includegraphics[width=.4\textwidth]{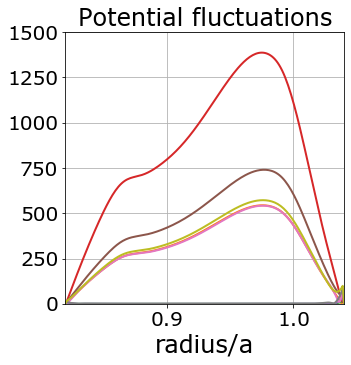} &
\includegraphics[width=.4\textwidth]{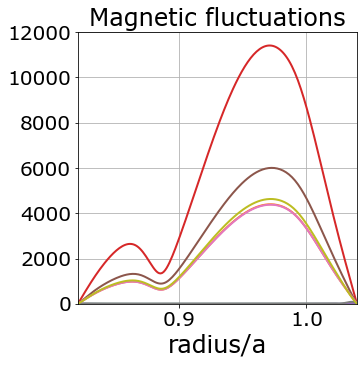} 
\end{tabular}
  \caption{Amplitude of the pressure (left), electro-static potential (center) and $\psi$ (right) fields for the 1D model before and during the pressure growing phase (orange, pink, green and purple), at the top (grey) and during the relaxation (red, brown and yellow). The curves correspond to the orange, green, red, purple, brown, pink, grey and yellow points in figure \ref{fig_1D2}.}
\label{fig_1D_flu}
\end{figure}

In what follows, it has been considered the mode with $m=11$ and $n=4$. In figures \ref{fig_1D2}, \ref{fig_1D_prq} and \ref{fig_1D_flu} the results of the simulations of the 1D model are presented.    

The average pressure is plotted in figure \ref{fig_1D2} and it exhibits the same kind of spikes as in 3D simulations, even though the oscillations are less pronounced. This achievement confirms that the $E\times B$ sheared mean flow, that is neglected in the 1D model, does not play an active role in relaxations.

The profiles of the 1D model fields are shown in figures \ref{fig_1D_prq} and \ref{fig_1D_flu} at some specific times, corresponding to both the pressure growing phase and the relaxation. These profiles confirm the expectations from 3D simulations: during the pressure growing phase (green, purple and grey curves) the pressure gradients steepen and the fluctuations are low, while during relaxation (red, brown and yellow curves) pressure collapses and fluctuations are large. In particular, by comparing the grey and red curves one sees how fluctuations have a bump at the beginning of the relaxation, while later they get reduced (brown and yellow curves) and return back to the amplitudes they had before the pressure growing phase (orange and pink curves). This is the same qualitative behavior shown in figure \ref{fig_flu}. Quantitatively, the 1D model is more unstable with respect to 3D simulations resulting in stronger reduction of the fluctuations during the pressure growing phase and more violent resurgence during relaxation. This feature can also explain why oscillation are less pronounced in the 1D model: fluctuations grow faster and back-react on the pressure profile sooner, implying a briefer pressure growing phase.  

\begin{figure}%[H]
\centering
\begin{tabular}{c}
\includegraphics[width=.8\textwidth]{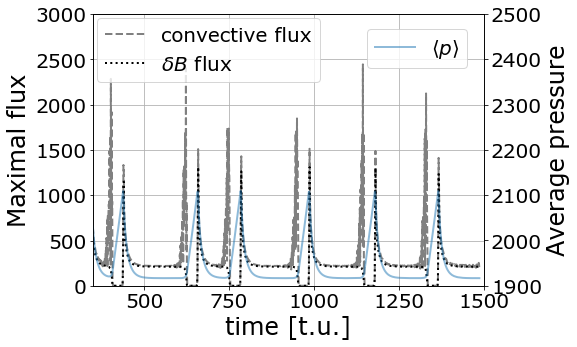} \\
\includegraphics[width=.8\textwidth]{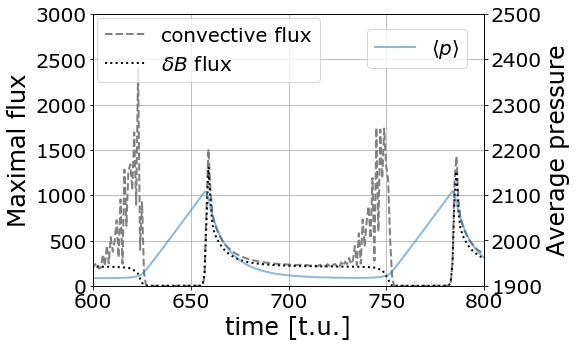} 
\end{tabular}
  \caption{Maximal convective (dashed, grey) and $\delta B$ (dotted, black) fluxes (left y-axis) and averaged pressure (blue, right axis) vs time for the 1D model and $Q=60$. The bottom figure is a zoom of the top one over the second and the third oscillation.}
\label{fig_1D1}
\end{figure}

The maximal convective and $\delta B$ fluxes are shown in figure \ref{fig_1D1} and they exhibits two kinds of peaks: a higher peak of the convective flux only, preceding the pressure growing phase, and a lower peak of both the convective and $\delta B$ fluxes at the maximal average pressure values. These lower peaks occur at the beginning of the relaxation in qualitative agreement with the results obtained for 3D simulations shown in figure \ref{fig3}. Moreover the magnitude of convective and magnetic turbulence is of the same order. %This fact confirms the validity of the choice made for $\alpha$, which has been amplified to enhance the effect of diamagnetic stabilization. 
 
The higher peaks are novel features of the 1D model. In 3D simulations the nonlinear interactions between the modes during relaxation provide a reduction of the fluctuations and of the convective flux which seeds a new pressure growing phase. In the 1D model, the nonlinear interaction between the modes are neglected, since a single mode is considered, and there is a bump of the convective flux just before a new pressure growing phase. Henceforth, in the 1D model convection plays the role of the nonlinear interaction in reducing the fluctuations and setting up the conditions for a new oscillation.    

\section{Conclusions}
A novel class of oscillations has been derived from EMEDGE3D simulations in the presence of the diamagnetic coupling. These oscillations are characterized by a strong dependence from the plasma beta parameter $\beta_e$ within the Ohm's law, with lower $\beta_e$ providing larger amplitude, and have been described in terms of a linear increase in time of the average pressure followed by a sudden drop. Drift-wave modes are the relevant modes that are stabilized during the pressure growing phase and back-react on the equilibrium pressure profile to induce the relaxation. This scenario has been confirmed by inspecting the equilibrium and fluctuation profiles, the wave-number/phase shift plot and  the nonlinear (convective and $\delta B$) fluxes. The latter, together with the energy flux into the scrape of layer, are negligible everywhere except at the beginning of the relaxation. The observed oscillations have the following similarities with type III ELMs% suggest that the observed oscillations are indeed type III ELMs 
\begin{itemize}
\item the energy flux into the scrape of layer decreases with increasing oscillation frequency, 
\item the oscillations are tamed by increasing plasma temperature.
\end{itemize}
This achievement can shed a new light on the determination of the physical conditions to generate transport barriers and to make them relax in a controlled and beneficial way.  

In this respect, the 1D model, that has been here constructed and showed to reproduce qualitatively the outcomes of 3D simulations, provides a simplified arena to test the mechanism behind the generation of relaxation oscillations.% type III ELMs. 

\section*{Acknowledgments}
One of the authors, F.C., benefited from co-financing by Aix-Marseille
University (AMU) and the French National Centre for Scientific Research
(CNRS).

\end{document}